\documentclass[aps,prl,amsmath,amssymb,twocolumn]{revtex4-2}
\usepackage[english]{babel}
\usepackage[utf8]{inputenc}
\usepackage{amsfonts}
\usepackage[T1]{fontenc}
\usepackage[pdftex]{graphicx}
\usepackage{fancyhdr}
\usepackage{hyperref}
\usepackage{url}
\usepackage[titletoc,title]{appendix}
\usepackage{epstopdf}
\usepackage{amsmath}
\usepackage{amssymb}
\usepackage{dsfont}
\usepackage{mathtools}
\usepackage{bbold}
\usepackage{xcolor}
\usepackage{tikz}
\usepackage{cancel}
\usetikzlibrary{arrows.meta}

\hypersetup{
    bookmarks=false,         
    unicode=false,          
    pdftoolbar=false,        
    pdfmenubar=true,        
    pdffitwindow=false,     
    pdfstartview={FitH},    
    pdftitle={},    
    pdfauthor={},     
    pdfsubject={},   
    pdfcreator={},   
    pdfproducer={}, 
    pdfkeywords={},
    pdfnewwindow=true,      
    colorlinks=true,       
    linkcolor=black,          
    citecolor=blue,        
    filecolor=magenta,      
    urlcolor=blue           
}

\graphicspath{{./Figures/}}

\newcommand{\ket}[1]{|#1\rangle}
\newcommand{\bra}[1]{\langle#1|}

\newcommand{\A}[1]{\hat{a}_{#1}}
\newcommand{\Adag}[1]{\hat{a}^\dagger_{#1}}

\begin{document}

\title{Non-reciprocal dynamics and the non-Hermitian skin effect of repulsively bound pairs}

\author{Pietro Brighi}
\affiliation{Faculty of Physics, University of Vienna, Boltzmanngasse 5, 1090 Vienna, Austria}
\author{Andreas Nunnenkamp}
\affiliation{Faculty of Physics, University of Vienna, Boltzmanngasse 5, 1090 Vienna, Austria}
\date{\today}

\begin{abstract}
    We study the dynamics of a Bose-Hubbard model coupled to an engineered environment which in the non-interacting limit induces an effective non-reciprocal hopping as described by the Hatano-Nelson model.
    At strong interactions, two bosons occupying the same site form a so-called repulsively bound pair, or doublon.
    Using tensor-network simulations, we clearly identify a distinct doublon lightcone and show that the doublon inherits non-reciprocity from that of single particles.
    Applying the idea of reservoir engineering at the level of doublons, we introduce a new set of dissipators and we analytically show that then the doublon dynamics are  governed by the Hatano-Nelson model.
    This brings about a \textit{two-particle} non-Hermitian skin effect and non-reciprocal doublon motion.
    Combining features of the two models we study, we show that single particles and doublons can be made to spread with opposite directionality, opening intriguing possibilities for the study of dynamics in interacting non-reciprocal models.
\end{abstract}

\maketitle

\begin{figure}[t]
    \includegraphics[width=.99\columnwidth]{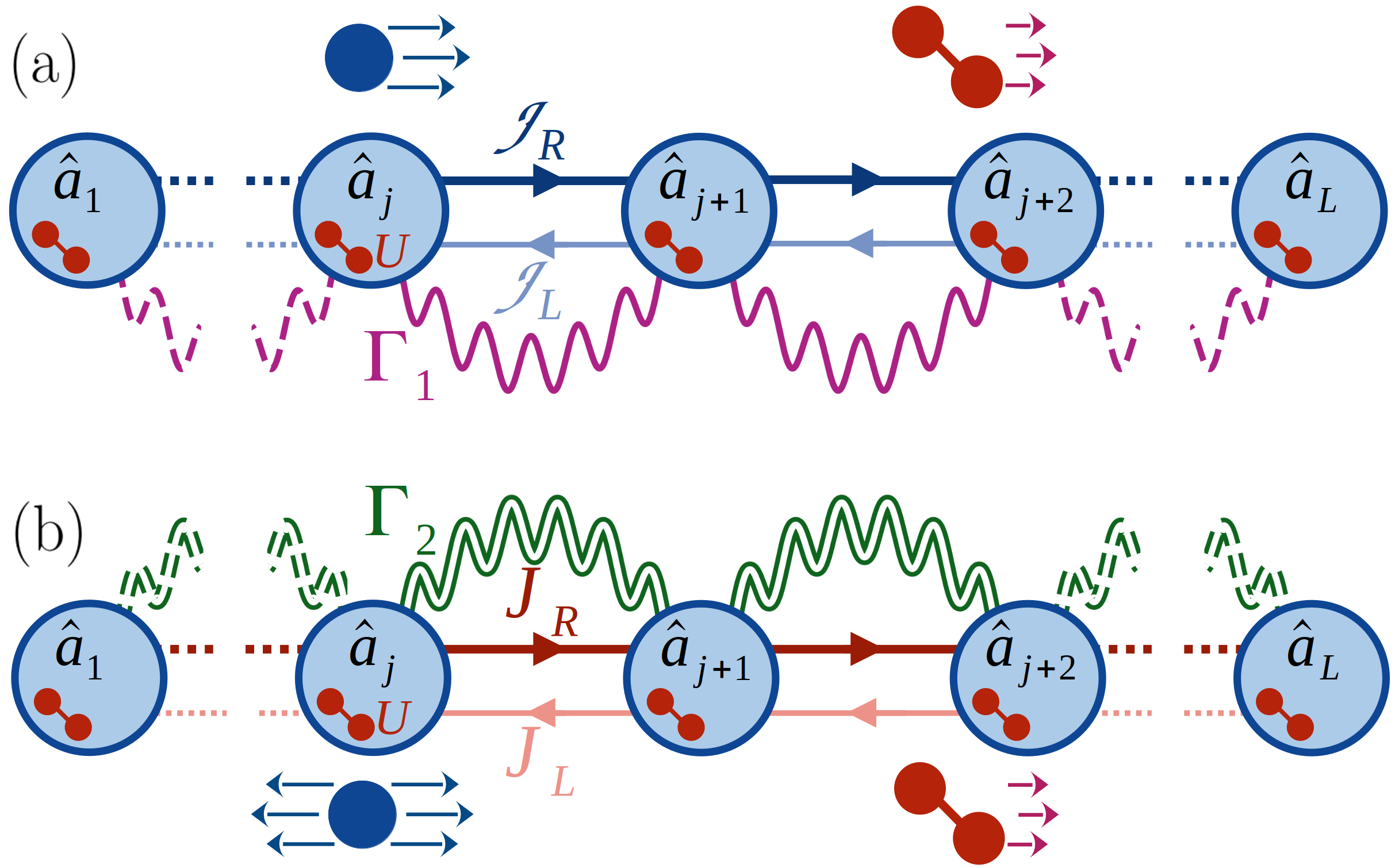}
    \caption{
    \label{Fig:Sketch}
    (a) Neighboring cavities are coupled with amplitude $J$ and bosons on the same site interact via the Kerr non-linearity with strength $U$. 
    In addition, the one-particle nearest-neighbor dissipation $\Gamma_1$ couples neighboring cavities [$\hat{L}_j$ in Eq.~(\ref{Eq:Lindblad})], giving rise to non-reciprocal hopping $\mathcal{J}_{R(L)}$.
    Both single-particle and doublon dynamics are directional, although doublons propagate more slowly and decay faster.
    (b) As in the strong coupling limit $U\gg J$ doublons are stable excitations, we introduce a second two-particle dissipator with rate $\Gamma_2$ [Eq.~(\ref{Eq:Doublon Diss})].
    In this second model, doublons acquire directional dynamics with non-reciprocal hopping $J_{R(L)}$, whereas single particles spread reciprocally.
    }
\end{figure}

\textit{Introduction -- }%
Non-reciprocal systems appear in physics in many different forms (see e.g.~\cite{Vitelli2021} and references therein). 
They have recently received significant attention for the interesting phenomena they host, often lacking a counterpart in reciprocal systems.
These range from the dramatic sensitivity of the spectral properties on the boundary conditions, to novel topological features in the complex spectrum~\cite{Ueda2018,Ueda2019a,Ueda2020,Bergholtz2021}.
A paramount example is the celebrated Hatano-Nelson model~\cite{Hatano1996} where the presence of non-reciprocal hopping leads to an exponential localization of all left and right eigenstates on the opposite boundaries of an open one-dimensional chain, the so-called non-Hermitian skin effect (NHSE)~\cite{Lee2016,Torres2018,Wang2018,Chen2022,Li2023}.

One physical implementation of non-reciprocity is based on reservoir engineering~\cite{Zoller1996,Clerk2015} where the system is coupled to a non-trivial environment resulting in an effective non-reciprocal dynamics.
These strategies are widely used in the non-interacting case~\cite{Tureci2018,Zhong2019,Porras2019,Wanjura2020,Nunnenkamp2021,Clerk2022,Clerk2023,Nunnenkamp2023,Rabl2023,Rabl2024} where the equations of motion close and one can analytically obtain the non-Hermitian Hamiltonian giving rise to non-reciprocity.
Additionally, reservoir engineering and non-reciprocity were recently realized in cold atoms experiments~\cite{Takahashi2017,Yan2020,Yan2022}.

The rich landscape of novel phenomena emerging in non-reciprocal systems has further stimulated the question of the interplay of non-reciprocity and interactions.
Most of the literature on non-Hermitian interacting systems focuses on the no-click limit of engineered open many-body systems~\cite{Ueda2018a,Ueda2019,Kawakami2019,Zhu2020,Chen2020,Chang2021,Ryu2022,Neupert2022,Brzezinska2022,Liang2022,He2022,Zhang2022,Pan2023,Park2023,longhi2023,Kawabata2023,Sun2023,Kou2023,Lee2023,Kawabata2024,Chen2024,Yoshida2024,Li2024,Zhou2024}.
This approximation, however, completely neglects the effect of jump operators and describes a single trajectory (the one where no photon is detected, hence the name) out of the exponentially many possibilities.
In certain particular cases, however, a formal relation between the eigensystem of the non-Hermitian Hamiltonian and the eigensystem of the Liouvillian can be established~\cite{Torres2014}.
A different possible step towards interacting systems considers the interaction on the mean-field level, yielding tractability at the expenses of a correct quantification of quantum fluctuations~\cite{Ezawa2022}.

To correctly take fluctuations into account, one can study the system at the level of the many-body Lindblad master equation~\cite{Wang2019,Ueda2021,Ueda2021a,Bergholtz2022,Bergholtz2024}.
While in this framework analytical and exact numerical methods have limited application, tensor network techniques allow the study of dynamics of large interacting systems.
The use of these methods is well established in the context of open quantum dynamics~\cite{Cirac2004,Prosen2009,Clark2010,Montangero2018}, and it was recently applied to the study of universality in a non-reciprocal spin $1/2$ $XXZ$ chain~\cite{Hanai2023}.

\begin{figure*}
    \centering
    \includegraphics[width=.99\textwidth]{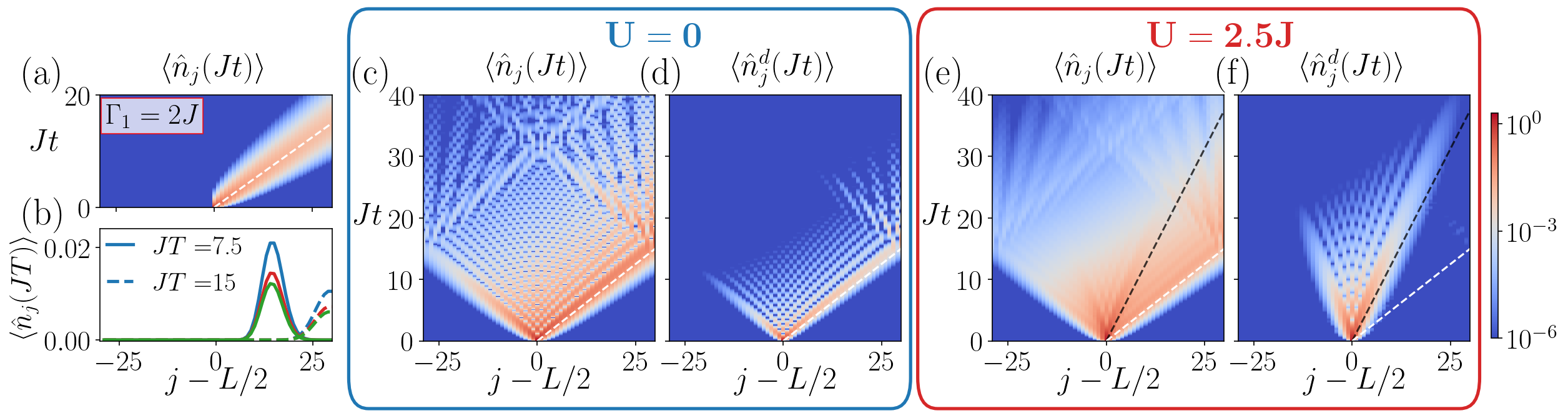}
    \caption{
    \label{Fig:nt}
    (a) At $\Gamma_1 = 2J$ , dynamics are restricted to the right half of the system and follow the single-particle lightcone (white dashed line) for all $U$ ($U=0$ is shown here).
    (b) Remarkably, the full non-reciprocity expected in the non-interacting case is extended to $U\neq0$ substantially unchanged (red and green lines), as shown by the snapshots at different times.
    (c)-(f) At smaller $\Gamma_1=0.1J$ the system is effectively less non-reciprocal, the left half of the system becomes slightly populated, and novel interaction-driven phenomena emerge.
    (e)-(f) In particular, at $U=2.5J$ we observe the appearance of a \textit{lightcone} corresponding to stable doublons moving non-reciprocally (black dashed line) and distinct from the single-particle lightcone observed in panels (c) and (d).
    This feature is more evident in the doublon density dynamics for $U=2.5J$ (f).
    The single-particle lightcone is highly suppressed, and the density propagation follows the black dashed line corresponding to $J^2/U$.
    These data were obtained for a system of $L=60$ sites, using an MPDO of bond dimension $\chi=128$ and $\theta_1=-\pi/2$.
    }
\end{figure*}

Here, we study the dynamics of a Bose-Hubbard model in presence of engineered dissipation.
Inspired by the so-called \textit{repulsively bound pairs} appearing in isolated optical lattices at strong interactions~\cite{zoller2006}, we focus on the dynamics of a single \textit{doublon}, i.e.~a composite particle made of two bosons occupying the same site.
Focusing on dynamics, our work unveils interesting features beyond the steady state which in the strong non-reciprocity regime is only weakly affected by the presence of interactions ~\cite{Hanai2023}.
Our numerical simulations show that both single particles and doublons can move non-reciprocally, although with different velocities.
The directional doublon lightcone we identify is a clear indication of emergent non-reciprocity in the interacting regime.

Applying the idea of reservoir engineering~\cite{Clerk2015} to the effective strong-coupling doublon Hamiltonian, we introduce a new set of dissipators and show that the resulting equations of motion within the single-doublon sector reproduce the Hatano-Nelson model.
Our model is then characterized by an \textit{interaction-induced} NHSE, where single-particle dynamics are reciprocal and \textit{only doublons} feature directional motion.
Their different behavior opens several intriguing possibilities for the study of dynamics, as we show by briefly exploring the case of opposite directionality for doublons and single particles.

\textit{Model -- }%
We study bosons in a dissipative cavity array with on-site Kerr non-linearity of strength $U$. 
The coherent part of the dynamics is encoded in the Bose-Hubbard Hamiltonian 
\begin{equation}
    \label{Eq:H}
    \begin{split}
    \hat{H} &= J\sum_{j=1}^{L-1} (\Adag{j}\A{j+1} + \text{H.c.}) + \sum_{j=1}^L U (\Adag{j})^2\A{j}^2,
    \end{split}
\end{equation}
where $J$ is the hopping amplitude between neighboring cavities and $\Adag{j}$($\A{j}$) creates(annihilates) a boson on site $j$.
The action of the environment is represented by a set of Lindblad operators $\hat{L}_j$ which introduce a nearest-neighbor dissipation $\hat{L}_j = \sqrt{\Gamma_1}(\A{j} + e^{\imath\theta_1}\A{j+1})$.
The full quantum dynamics of the system are then described by the many-body Lindblad master equation
\begin{equation}
    \label{Eq:Lindblad}
        \dot{\rho} =  -\imath[\hat{H},\rho] + \sum_{j=1}^{L-1} \hat{L}_j\rho \hat{L}_j^\dagger - \frac{1}{2}\{\hat{L}_j^\dagger\hat{L}_j,\rho\},
\end{equation}
as depicted in Figure~\ref{Fig:Sketch}(a).
This model was recently introduced in Refs.~\cite{Chen2024,Bergholtz2024} studying the effective non-Hermitian Hamiltonian arising in its fully non-reciprocal no-click limit.

Previous studies in the non-interacting case, $U=0$, have shown how the dynamics of the first moments $\langle \A{j}\rangle$~\cite{Wanjura2020} and the second moments $\langle \Adag{j}\A{i}\rangle$~\cite{Clerk2022} are described by a non-Hermitian dynamical matrix corresponding to the Hatano-Nelson model~\cite{Hatano1996}.
Consequences of this effective non-reciprocity include directional exponential amplification of the cavity amplitude~\cite{Porras2019,Wanjura2020,Wanjura2021} and non-reciprocal dynamics of the boson densities~\cite{Clerk2023}.

\begin{figure*}
    \centering
    \includegraphics[width = .99\textwidth]{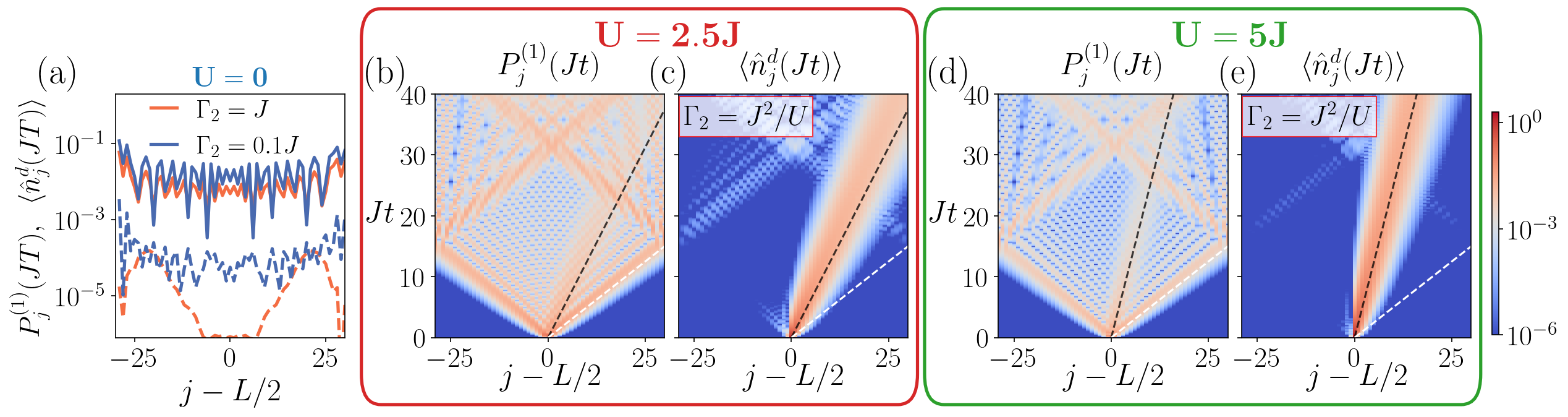}\\
    \caption{\label{Fig:density quadratic diss}
    (a) In absence of interactions, $U=0$,  dynamics are  reciprocal, irrespective of dissipation rate $\Gamma_2$.
    Both single-particle (solid lines) and doublon (dashed lines) population profiles at time $JT=15$ are equally spread in the two halves of the system.
    As expected, the value of $\Gamma_2$ affects significantly the doublon density only.
    (b)-(e) As interactions are turned on $U\gg J$ the system shows clear signatures of directional motion.
    (c),(e) Due to the destructive interference induced by the quadratic dissipator, a highly non-reciprocal doublon lightcone $\tilde{x}\propto J^2/U$ (black dashed line) appears, affecting the doublon density.
    (b),(d) The single-particle population, however, spreads reciprocally and is only slightly affected by the doublon lightcone due to its decay into single-particle states.
    (e),(d) As the interaction strength is increased, the non-reciprocal doublon spreading becomes slower, in agreement with the smaller effective hopping amplitude.
    Additionally, we notice a smaller population of single-particle states due to the increased stability of the doublon.
    These data were obtained for a system of $L=60$ sites, using an MPDO of bond dimension $\chi=128$ and $\theta_2=-\pi/2$.
    }
\end{figure*}

\textit{Non-reciprocity and dynamics -- }%
To avoid the exponential growth of the Hilbert space ($\mathcal{D} \approx 10^{6}$ in the case we study), we use tensor-network methods which allow to obtain accurate results at a cost scaling only linearly with system size.
In particular, we use the well-known time-evolving block decimation (TEBD) algorithm~\cite{Vidal2003} adapted to the description of Lindblad dynamics, as detailed in the Supplementary Material~\cite{Supplementary}.

Throughout this work, we study the dynamics of the single-doublon initial state
\begin{equation}
    \label{Eq:psi0}
    \ket{\psi_0} = \frac{1}{\sqrt{2}}(\Adag{L/2})^2\ket{\text{vac}},
\end{equation}
which in the isolated scenario can form a \textit{repulsively bound pair}~\cite{zoller2006} when strong interactions make single-particle hopping energetically unfavorable.
The behavior of this composite particle at $U\gg J$ is accurately captured by the following effective Hamiltonian~\cite{Lukin2003,Svistunov2003} which can be obtained through a Schrieffer-Wolff transformation~\cite{SW,Divincenzo2011}
\begin{equation}
    \label{Eq:Heff}
    \hat{H}_\text{eff} = \frac{1}{2}\frac{J^2}{U}\sum_{j=1}^{L-1} \left[(\Adag{j})^2(\A{j+1})^2 + \text{H.c.}\right],
\end{equation}
where doublons move coherently with a reduced hopping amplitude.
The presence of engineered dissipation makes the dynamics richer as it enables non-reciprocal hopping in the non-interacting case $\mathcal{J}_{R(L)} = J-\imath e^{-(+)\imath\theta_1}\Gamma_1/2$, tuning the model from  reciprocal at $\Gamma_1 = 0$ to fully non-reciprocal at $\Gamma_1=2J$ (for $\theta_1 = \pm \pi/2$)~\cite{Clerk2015}.

Considering second-order processes in the equations of motion of doublon states, one can show that doublons inherit non-reciprocity and move with an effective non-reciprocal hopping amplitude $\mathcal{J}^{(d)}_{R(L)}\approx \mathcal{J}^2_{R(L)}/(U +\imath \Gamma_1)$.
We confirm this prediction numerically in Figure~\ref{Fig:nt}.

At strong non-reciprocity $\Gamma_1=2J$ and $\theta_1 = -\pi/2$ (a)-(b), particles move only to the right following the single-particle \textit{lightcone} $x(t)\propto J t$, irrespective of interaction strength $U$.
This is a consequence of the large dissipation rate $\Gamma_1$, which quickly depletes the system thus making the effect of interactions weak~\cite{Hanai2023}.

In Figure~\ref{Fig:nt} (c)-(f) we decrease the dissipation rate to $\Gamma_1 = 0.1J$, hence the degree of non-reciprocity is expected to be weaker.
Nevertheless, dynamics still show clear signatures of non-reciprocity in both the interacting and non-interacting cases.
Importantly, the presence of interactions leads to the emergence of a \textit{second lightcone} $\tilde{x}(t) \propto (J^2/U) t$ [black dashed line in panels (e) and (f)].
This doublon lightcone is related, to leading order, to the effective non-reciprocal doublon hopping amplitude, and it clearly highlights the slower doublon dynamics due to strong non-linearity.
We further observe a slightly larger population on the right branch of the doublon lightcone, suggesting the extension of non-reciprocity also to the interacting level.
In passing, comparing panels (c) and (e) we notice that the interference pattern at long times $Jt\gg L/2$ is washed out in the interacting case, reminiscent of many-body dephasing in isolated systems, where dynamics relax to the thermal average due to interactions and ergodicity~\cite{Rigol2008,Vengalattore2011,Izraeliev2012,d'alessio2016}.

To obtain a clearer picture of the doublon dynamics and distinguish it from that of single particles it is useful to define a \textit{doublon density}
\begin{equation}
    \label{Eq:nd}
    \hat{n}^d_j = \frac{1}{2}(\Adag{j})^2\A{j}^2
\end{equation}
which is identically zero for all single-particle states and in the single-doublon sector corresponds to the doublon population on site $j$.
In Figure~\ref{Fig:nt} (d) and (f) we show the dynamics of $\langle \hat{n}^d_j\rangle$ for a small dissipation rate $\Gamma_1 = 0.1J$ and for $U=0$ and $U=2.5J$, respectively.
In the non-interacting case, the doublon density follows the single-particle lightcone, indicating the absence of coherent doublon motion.
On the other hand, at $U=2.5J$, the single-particle lightcone is strongly suppressed and the dominant contribution comes from the slower and non-reciprocal doublon motion.
Hence, the interacting system inherits the non-reciprocity characterizing single-particle dynamics.
The emergence of metastable non-reciprocal doublons is a genuine consequence of interactions, clearly distinguishable from the single-particle case, and is one of the main results of this letter.

\textit{Stabilizing doublon directional motion -- }%
Inspired by the structure of the effective Hamiltonian Eq.~(\ref{Eq:Heff}), we introduce a different set of dissipators which \textit{stabilize} doublon non-reciprocity.
Following the approach of Ref.~\cite{Clerk2015}, we replace the one-particle nearest-neighbor dissipator with its doublon version
\begin{equation}
    \label{Eq:Doublon Diss}
    \Gamma_2\mathcal{D}[\A{j}^2 + e^{\imath\theta_2}\A{j+1}^2],
\end{equation}
where only pairs of bosons (i.e.~doublons) are lost to the environment, as sketched in Figure~\ref{Fig:Sketch} (b).

Using the effective Hamiltonian~(\ref{Eq:Heff}) and the dissipator above we obtain the equation of motion for the doublon amplitudes $\langle\A{\ell}^2\rangle$ and correlations $\langle (\Adag{\ell})^2\A{m}^2\rangle$~\cite{Supplementary}.
In the single-doublon sector these simplify and can be written in terms of a non-Hermitian dynamical matrix $\mathbb{H}$ acting non-trivially on the \textit{single-doublon} space only
\begin{align}
    \label{Eq:EOM doublon n}
        \imath \frac{\partial\langle (\Adag{\ell})^2\A{m}^2\rangle}{\partial t} &= \sum_j \mathbb{H}_{m,j} \langle (\Adag{\ell})^2\A{j}^2\rangle - \mathbb{H}^\dagger_{j,\ell}\langle (\Adag{j})^2\A{m}^2\rangle,
\end{align}
with $\mathbb{H}$ the Hatano-Nelson matrix
\begin{equation}
    \label{Eq:H matrix}
    \mathbb{H} = \sum_j J_R \ket{j+1}_2\bra{j}_2 + J_L\ket{j-1}_2\bra{j}_2 -2\imath \Gamma_2 \ket{j}_2\bra{j}_2.
\end{equation}
Similarly to the non-interacting case~\cite{Wanjura2020,Clerk2022}, the interference of the coherent nearest-neighbor coupling with the doublon dissipation causes the emergence of different left and right hopping amplitudes $J_L = \frac{J^2}{U} - \imath e^{\imath\theta_2}\Gamma_2 $ and $J_R = \frac{J^2}{U} - \imath e^{-\imath\theta_2}\Gamma_2 $.
The dissipation rate required for full non-reciprocity is then $\Gamma_2=J^2/U$ and is of the same order of the doublon motion timescale.
This results in more stable non-reciprocal doublon dynamics, as compared to the one-particle dissipator case where full non-reciprocity is achieved at $\Gamma_1=2J\gg J^2/U$.

Remarkably, the non-Hermitian skin effect arising from the Hatano-Nelson matrix~(\ref{Eq:H matrix}) affects only doublon states $\ket{j}_2 = \ket{0\dots 2_j\dots 0}$.
Binding particles together, interactions have a dramatic effect and enable the exponential localization of doublons at the boundaries of the system, whereas single-particle behave reciprocally.
The non-reciprocity and non-Hermitian skin effect resulting from the quadratic dissipator~(\ref{Eq:Doublon Diss}) combined with the interaction-induced doublon stability represent the second central result of our work.

In our numerical analysis, we go beyond the approximate picture of the effective Hamiltonian and simulate the full dynamics of the system using the interacting Hamiltonian~(\ref{Eq:H}) and the quadratic dissipator~(\ref{Eq:Doublon Diss}).
As we want to separate single-particle from doublon contributions to the dynamics, the density $\hat{n}_j = \Adag{j}\A{j}$ is not a convenient quantity, as it is affected by both.
We then define the single-particle population $\hat{P}^{(1)}_j = \hat{n}_j - 2\hat{n}^d_j$ which accounts for the weight of single-particle states, when the total number of bosons is $N=2$.

\begin{figure}[t]
    \centering
    \includegraphics[width=.99\columnwidth]{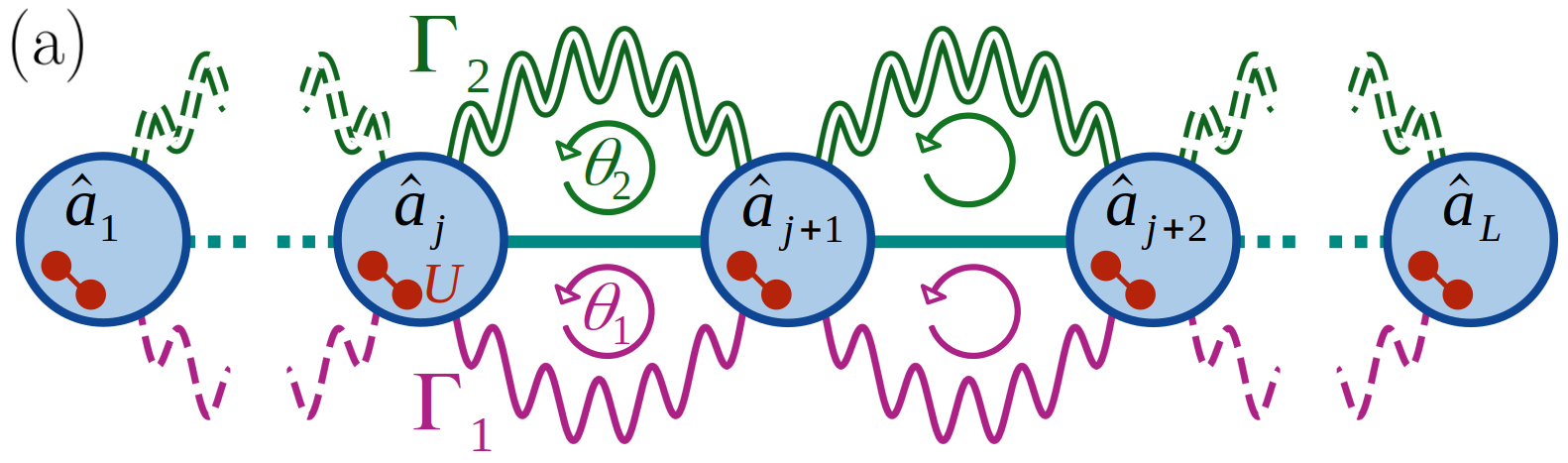} \\
    \includegraphics[width=.99\columnwidth]{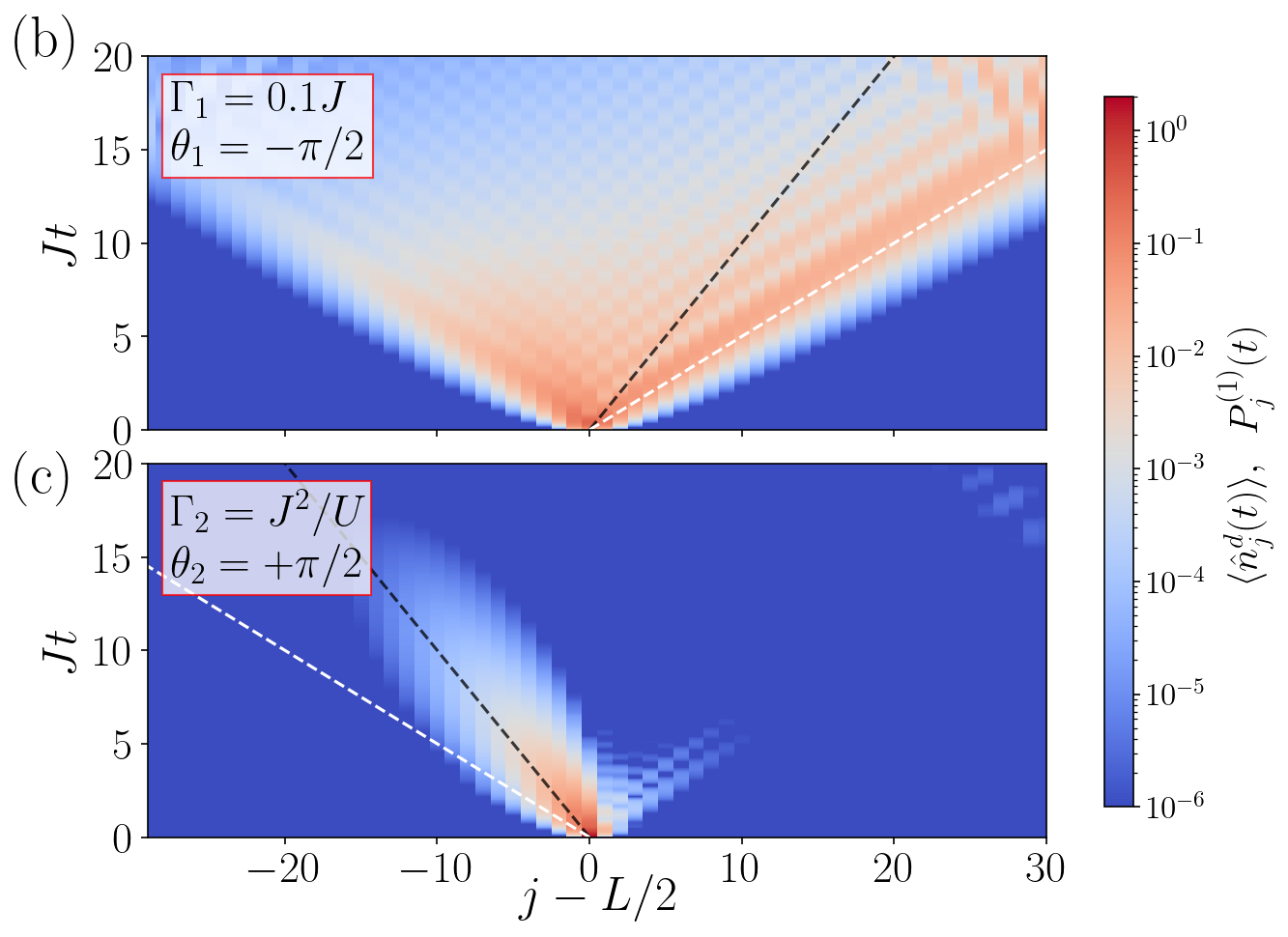}
    \caption{
    (a) Combining the one- and two-particle nearest-neighbor dissipators gives rise to fascinating effects on particle dynamics.
    (b) At $\theta_1 = -\pi/2$ and $\Gamma_1 = 0.1J$ the single-particle density is slightly non-reciprocal towards the \textit{right}.
    (c) However, choosing $\theta_2 = +\pi/2$ and $\Gamma_2 = J^2/U$ leads to almost full non-reciprocity of doublons to the \textit{left}.
    These data were obtained for a system of $L=60$ sites, using an MPDO of bond dimension $\chi=128$ and $U=2J$.
    \label{Fig:combined diss}
    }
\end{figure}

In Figure~\ref{Fig:density quadratic diss} (a), we show the single-particle and doublon profiles at time $JT=15$ in the non-interacting case.
Due to the absence of stable doublons at $U=0$, dynamics are reciprocal and particles spread in both directions equally, irrespective of the value of $\Gamma_2$.

A dramatic difference is observed when $U\gg J$ [Figure~\ref{Fig:density quadratic diss}(b)-(e)], where the doublon forms a stable excitation and the quadratic dissipative coupling~(\ref{Eq:Doublon Diss}) leads to strong non-reciprocity.
In particular, at $\Gamma_2 = J^2/U$ and $\theta_2=-\pi/2$, where $J_L = 0$, the doublon density propagates exclusively towards the right boundary as predicted by the equations of motion~(\ref{Eq:EOM doublon n}).
As a consequence of the finite interaction $U$, single-particle hopping processes are allowed, and the initial doublon can decay into single-particle states.
These are free to propagate in both directions, as they are not affected by the dissipator~(\ref{Eq:Doublon Diss}).

\textit{Opposite directionality -- }
Combining the one-particle and doublon nearest-neighbor dissipators as in Figure~\ref{Fig:combined diss} (a), one can separately control doublons and single particles.
In Figure~\ref{Fig:combined diss}, we show dynamics of both single-particle population (b) and doublon density (c).
Choosing $\theta_1 = -\theta_2$ generates opposite interference of the nearest-neighbor hopping with the two dissipators and results in the different directionality of single particles (b) and doublons (c).
This simple example points out how introducing the dissipator~(\ref{Eq:Doublon Diss}) causing \textit{doublon non-reciprocity} opens several interesting directions for the investigation of interacting non-reciprocal dynamics.

\textit{Conclusions -- }%
In this work, we investigated the dynamics of a Bose-Hubbard model coupled to engineered dissipators.
Showing the emergence of a non-reciprocal doublon lightcone, we highlighted how the study of time evolution can unveil genuinely interacting effects, which would be hidden in the study of steady-states alone~\cite{Hanai2023}.

We introduced a novel type of dissipator, based on the structure of the strongly-interacting effective Hamiltonian, and showed how it gives rise to a doublon non-Hermitian skin effect.
This arises at the level of the Lindblad master equation, going beyond the no-click limit studied in previous works~\cite{Yoshida2024,Li2024}.
The quadratic dissipator leads to the two-particle non-reciprocal dynamics observed in our numerical simulations, and opens new possibilities for the study of dynamics in non-reciprocal systems.

Strictly related to the study of dynamics we presented is the issue of how relaxation and thermalization are affected by interactions in non-reciprocal models~\cite{Ueda2021,Clerk2023}.
Our approach can be easily generalized to other systems which support stable excitations such as one-dimensional spin $1/2$ chains.
This setup allows for the study of strongly-correlated systems, raising some intriguing questions regarding non-Hermitian topology of many-body systems~\cite{Ryu2022} as well as the nature of transport in non-reciprocal bosonic and fermionic models~\cite{Rabl2023,Rabl2024}.

\begin{acknowledgments}
\textit{Acknowledgments -- }%
P.~B.~is supported by the Erwin Schr\"odinger center for Quantum Science \& Technology (ESQ) of the \"Osterreichische Akademie der Wissenschaften (\"OAW) under the  Discovery Grant.
This work has been supported by Austrian Science Fund (FWF): COE 1 Quantum Science Austria.
The numerical simulations were performed using the ITensor library~\cite{ITensor} on the Vienna Scientific Cluster (VSC).
\end{acknowledgments}

\clearpage
\newpage
\onecolumngrid
\appendix
\begin{center}
    \textbf{Supplementary Material for ``Non-reciprocal dynamics and the non-Hermitian skin effect of repulsively bound pairs"}
    
    Pietro Brighi and Andreas Nunnenkamp

    \small{\textit{Faculty of Physics, University of Vienna, Boltzmanngasse 5, 1090 Vienna, Austria}}
\end{center}

\renewcommand{\theequation}{S\arabic{equation}}
\setcounter{equation}{0}
\renewcommand{\thefigure}{S\arabic{figure}}
\setcounter{figure}{0}
\section{Details on numerical methods}\label{Sec:App:Numerics}
\setcounter{page}{1}

In order to solve numerically the Lindblad master equation Eq.~(\ref{Eq:Lindblad}), one needs to \textit{vectorize} the density matrix $\rho$, replacing the density operator with a vector in a larger Hilbert space $\rho = \sum_{i,j}\rho_{ij}\ket{i}\bra{j} \to \sum_{i,j}\rho_{ij}\ket{j}\otimes\ket{i}$.
This convention allows to write the Liouvillian $\mathcal{L}$ as a $\mathcal{N}^2\times \mathcal{N}^2$ matrix, where $\mathcal{N}$ is the dimension of the original Hilbert space.
Once the problem is cast in terms of matrices and vectors, it is then straightforward to use conventional linear algebra numerical methods to obtain dynamics and steady state.
Due to the exponential growth of the Hilbert space, however, the exact numerical solution of the Lindblad master equation is a formidable task~\cite{Orus2021}.
Therefore, in our work we used tensor-network methods, which allow an accurate approximation of the dynamics of the system.

\begin{figure}[b]
    \centering
    \includegraphics[width=.32\columnwidth]{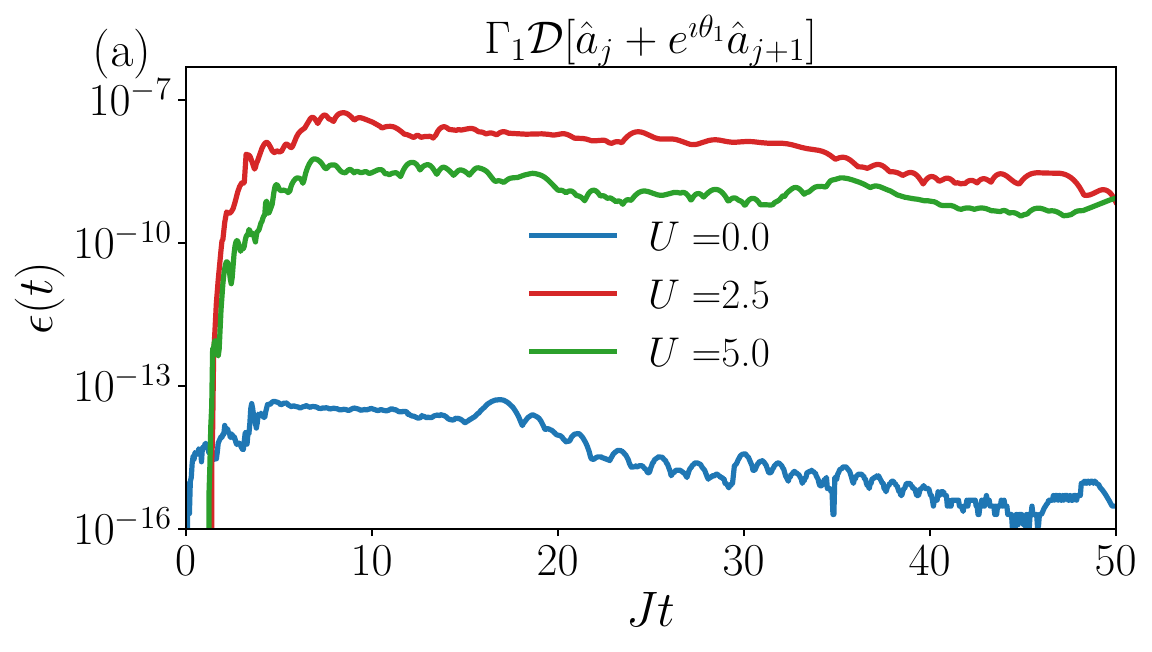}
    \includegraphics[width=.32\columnwidth]{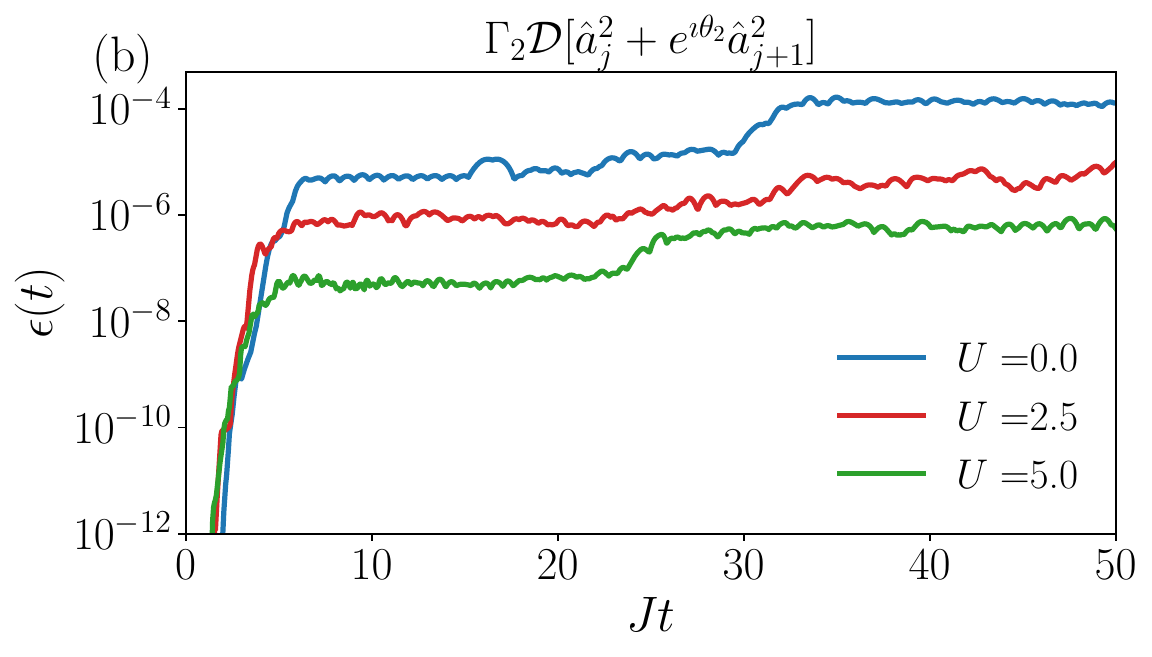}
    \includegraphics[width=.32\columnwidth]{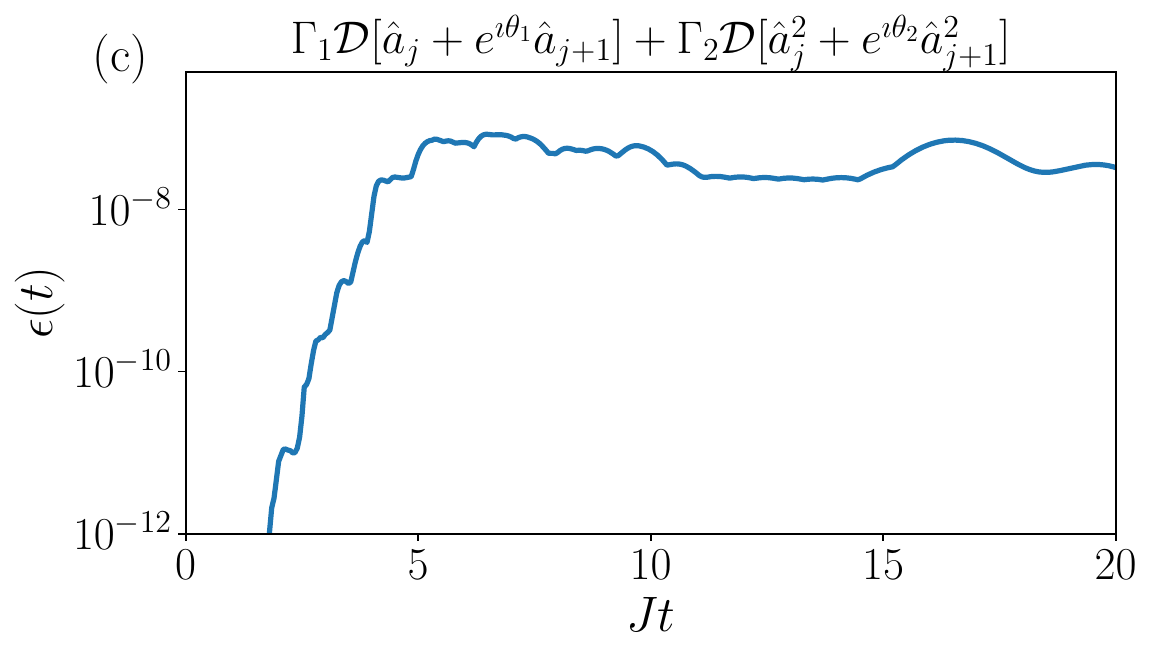}
    \caption{
    \label{Fig:relative error}
    The maximum error for all dissipators used in the main text remains very small, confirming the accuracy of the results presented.
    For the single-particle dissipator (a) we show $\Gamma_1=0.1J$ and $\theta_1 = -\pi/2$ as it represents the worse case for entanglement growth.
    In panel (b), instead, we show $\Gamma_2=J^2/U$ and $\theta_2=-\pi/2$, the values used in the main text.
    Finally, in panel (c) we show the error for the combined dissipators with ($\Gamma_1 = 0.1J,\theta_1=-\pi/2$) , ($\Gamma_2=J^2/U,\theta_2 = \pi/2$) and $U=2J$.
    In all cases the error is obtained comparing $\chi_1 = 64$ and $\chi_2 = 128$.
    }
\end{figure}

In the vectorized Hilbert space, the Liouvillian matrix describing the first model we introduce takes the following form
\begin{equation}
    \label{Eq:Lindblad matrix}
    \begin{split}
        \mathcal{L} &= -\imath(\mathbb{1}\otimes\hat{H} - \hat{H}^T\otimes\mathbb{1}) +  \Gamma_1 \sum_{j=1}^{L-1} (\Adag{j} + e^{-\imath\theta_1}\Adag{j+1})^T\otimes(\A{j} + e^{\imath\theta_1}\A{j+1}) \\
        &- \frac{1}{2}\Gamma_1 \sum_{j=1}^{L-1}\left\{\mathbb{1}\otimes(\Adag{j} + e^{-\imath\theta_1}\Adag{j+1})(\A{j} + e^{\imath\theta_1}\A{j+1}) +\left[(\Adag{j} + e^{-\imath\theta_1}\Adag{j+1})(\A{j} + e^{\imath\theta_1}\A{j+1})\right]^T\otimes\mathbb{1}\right\}.
    \end{split}
\end{equation}
Locality of both the Hamiltonian and the dissipators allows to efficiently write the Liouvillian as a low bond dimension matrix-product operator (MPO).
The vectorized density matrix can be similarly written in a matrix-product state (MPS) form.
To perform time evolution we use the well-known time evolving block decimation (TEBD) algorithm~\cite{Vidal2003} with local gates corresponding to local elements of the Liouvillian of Eq.~(\ref{Eq:Lindblad matrix}).

Throughout this work we used a $4$-th order Suzuki-Trotter decomposition with time-step $J\delta t = 0.05$.
First the Liouvillian is divided into its even and odd contributions
$$ \mathcal{L} = \sum_{i\in \text{even}}\mathcal{L}_{i,i+1} + \sum_{i\in \text{odd}}\mathcal{L}_{i,i+1}.$$
As even and odd terms do not commute, the propagator $e^{\mathcal{L}\delta t}$ cannot be written as a product of an even and an odd layer exactly.
One can however approximately do so, by applying a series of correcting layers which sum up to the correct time-step $\delta t$.
The error accumulated during this process is proportional to the order of the expansion $\mathcal{O}(\delta t^4)$~\cite{Suzuki2005}.
In our simulations, we take advantage of the fact that even and odd gates do commute among themselves (even with even and odd with odd) to perform a parallel evaluation of all the gates composing one layer.
Comparison with the exact dynamics on small systems $L=4$ and $N=2$ confirms the accuracy of TEBD, the only source of errors coming from trotterization.

\subsubsection{Convergence analysis}\label{Sec:App:bond dimension}

The time evolution will in general increase the bond dimension needed to accurately describe the full state.
The details on how fast and how much does entanglement grow, and therefore for how long is the simulation accurate, strongly depend on the initial state and on the amount of correlations developed during the dynamics.
As these are not known a priori, any simulation requires a convergence benchmark against different bond dimensions.

Therefore, we compare the expectation value of relevant observables obtained with different bond dimensions and define the maximum error in the array $\epsilon_n(\chi)$
$$
\epsilon_n(\chi,t) = \max_j{ |\langle\hat{n}^{\chi_1}_j(t)\rangle - \langle\hat{n}^{\chi_2}_j(t)\rangle|}
$$
where $\chi_1$ and $\chi_2$ are two different bond dimensions.

In Figure~\ref{Fig:relative error} we report the results of our convergence analysis for the different dissipators studied in the main text.
The maximum error at each time is calculated using $\chi_1=64$ and $\chi_2=128$.
The data confirm the accuracy of our simulations, as the maximum error is extremely small at all times.
This is not surprising, as the system is very dilute since only two particles are present at time $t=0$ and the environment does not add particles to the system.

In particular, in panel (a) we notice how for the non-interacting case the error is particularly small, as the two particles are weakly correlated.
In the quadratic dissipator case (b) this trend is inverted, as at $U=0$ particles are only weakly dissipated to the environment, thus effectively increasing correlations with respect to the interacting case.

\section{Equations of motion for the quadratic dissipator}\label{Sec:App:EOMs quadratic}

In the main text we report the equations of motion (EOM) for the doublon density in the single-doublon case and in absence of loss and gain.
Here, we provide details on the derivation and on the regime it is valid in.

The equation of motion of the expectation value of an operator $\hat{O}$ in the open setting is given by
\begin{equation}
    \label{Eq:EOM op}
    \frac{\partial \langle \hat{O} \rangle}{\partial t} = \imath \langle [\hat{H},\hat{O}] + \sum_{\alpha,j}(\hat{L}^{(\alpha)}_j)^\dagger \hat{O} \hat{L}^{(\alpha)}_j - \frac{1}{2}\{(\hat{L}^{(\alpha)}_j)^\dagger\hat{L}^{(\alpha)}_j,\hat{O}\}\rangle .
\end{equation}
In the case studied in the main text, we assume to be in the strong coupling regime $U\gg J$, where the effective Hamiltonian Eq.~(\ref{Eq:Heff}) accurately describes the closed system.
We can then use $\hat{H}_\text{eff}$ in Eq.~(\ref{Eq:EOM op}) together with the two-particle dissipator Eq.~(\ref{Eq:Doublon Diss}).
For the doublon amplitude $\langle \A{\ell}^2\rangle$, then
\begin{equation}
    \label{Eq:EOM ampl suppl}
    \begin{split}
        &\frac{\partial\langle\A{\ell}^2\rangle}{\partial_t} = \imath \frac{J^2}{2U}\sum_{j=1}^{L-1}\left\langle[(\Adag{j})^2\A{j+1}^2,\A{\ell}^2] + \langle[(\Adag{j+1})^2\A{j}^2,\A{\ell}^2]\right\rangle + \Gamma_2\sum_{j=1}^{L-1} \left\langle\left[(\Adag{j})^2 + e^{-\imath\theta_2}(\Adag{j+1})^2\right]\A{\ell}^2\left[\A{j}^2 + e^{\imath\theta_2}\A{j+1}^2\right]\right\rangle \\
        &-\frac{1}{2}\Gamma_2\sum_{j=1}^{L-1}\left\langle \left\{\left[(\Adag{j})^2 + e^{-\imath\theta_2}(\Adag{j+1})^2\right]\left[\A{j}^2 + e^{\imath\theta_2}\A{j+1}^2\right],\A{\ell}^2\right\}\right\rangle = -\imath \left(\frac{J^2}{U} - \imath\Gamma_2e^{\imath\theta_2}\right)\left\langle2\Adag{\ell}\A{\ell}\A{\ell+1}^2 + \A{\ell+1}^2\right\rangle \\
        &- \imath \left(\frac{J^2}{U} - \imath\Gamma_2e^{-\imath\theta_2}\right)\left\langle2\Adag{\ell}\A{\ell}\A{\ell-1}^2 + \A{\ell-1}^2\right\rangle + 2\Gamma_2\left\langle 2\Adag{\ell}\A{\ell}^3 +\A{\ell}^2\right\rangle.
    \end{split}
\end{equation}
The equation above simplifies in the particular case of a single doublon.
In fact, as the three-particle sector is never populated, the terms $\langle\Adag{\ell}\A{\ell}\A{\ell+1}^2\rangle$ and similar exactly vanish.
Therefore the simpler expression
\begin{equation}
        \imath \frac{\partial\langle\A{\ell}^2\rangle}{\partial_t} = J_R\langle\A{\ell+1}^2\rangle + J_L\langle\A{\ell-1}^2\rangle - 2 \imath\Gamma_2 \langle\A{\ell}^2\rangle
\end{equation}
follows, where $J_R = \left(\frac{J^2}{U} - \imath e^{-\imath\theta_2}\Gamma_2 \right) $ and $J_L = \left(\frac{J^2}{U} - \imath e^{\imath\theta_2}\Gamma_2\right) $.
The dynamics of the doublon amplitude in the strong coupling limit is hence determined by a non-Hermitian matrix identical to that of the Hatano-Nelson model.

Similarly, we obtain the equations of motion for the doublon correlations
\begin{equation}
    \begin{split}
    &\frac{\partial \langle (\Adag{\ell})^2\A{m}^2\rangle }{\partial_t} = \imath \frac{J^2}{2U}\sum_{j=1}^{L-1}\left\langle [(\Adag{j})^2\A{j+1}^2,(\Adag{\ell})^2\A{m}^2] + [(\Adag{j+1})^2\A{j}^2,(\Adag{\ell})^2\A{m}^2] \right\rangle \\
        &+ \Gamma_2\left[\sum_{j=1}^{L-1} \left\langle\left[(\Adag{j})^2 + e^{-\imath\theta_2}(\Adag{j+1})^2\right](\Adag{\ell})^2\A{m}^2\left[\A{j}^2 + e^{\imath\theta_2}\A{j+1}^2\right] \right\rangle - \frac{1}{2}\left\langle\left\{\left[(\Adag{j})^2 + e^{-\imath\theta_2}(\Adag{j+1})^2\right]\left[\A{j}^2 + e^{\imath\theta_2}\A{j+1}^2\right],(\Adag{\ell})^2\A{m}^2\right\}\right\rangle\right] \\
        &= -\imath J_R \left\langle2(\Adag{\ell})^2\Adag{m}\A{m}\A{m+1}^2 +(\Adag{\ell})^2\A{m+1}^2\right\rangle -\imath J_L \left\langle2(\Adag{\ell})^2\Adag{m}\A{m}\A{m-1}^2 +(\Adag{\ell})^2\A{m-1}^2\right\rangle  + \imath J^*_R \left\langle2(\Adag{\ell+1})^2\Adag{\ell}\A{\ell}\A{m}^2 +(\Adag{\ell+1})^2\A{m}^2\right\rangle\\
        &+ \imath J^*_L \left\langle2(\Adag{\ell-1})^2\Adag{\ell}\A{\ell}\A{m}^2 +(\Adag{\ell-1})^2\A{m}^2\right\rangle -4\Gamma_2\left\langle (\Adag{\ell})^2\Adag{\ell}\A{\ell}\A{m}^2 +(\Adag{\ell})^2\A{m}^2  + (\Adag{\ell})^2\Adag{m}\A{m}\A{m}^2\right\rangle
    \end{split}
\end{equation}
Again the equations of motion simplify once the single-doublon sector is considered.
Removing the vanishing terms, then, we obtain the following equation
\begin{equation}
    \begin{split}
    \imath &\frac{\partial \langle(\Adag{\ell})^2\A{m}^2\rangle}{\partial_t} = J_R\langle (\Adag{\ell})^2\A{m+1}^2\rangle + J_L\langle (\Adag{\ell})^2\A{m-1}^2 \rangle - J_L^*\langle (\Adag{\ell-1})^2\A{m}^2 \rangle - J_R^*\langle (\Adag{\ell+1})^2\A{m}^2 \rangle -4\imath\Gamma_2\langle(\Adag{\ell})^2\A{m}^2 \rangle.
    \end{split}
\end{equation}
These equations of motion can be conveniently expressed as the matrix-vector multiplication used in Eq.~(\ref{Eq:EOM doublon n}) in the main text.

\section{Physical implementation of the two-particle dissipator}\label{Sec:App:aux cavity}

The quadratic dissipator in the main text Eq.~(\ref{Eq:Doublon Diss}) can be obtained through the adiabatic elimination of an auxiliary strongly damped cavity mode $\hat{c}$.
The auxiliary cavity is coupled to the main array via the Hamiltonian 
\begin{equation}
    \hat{H}_c = J' \hat{c}^\dagger(\A{j}^2 + \A{j+1}^2) + \text{H.c}
\end{equation}
and it is strongly damped with a dissipation rate $\gamma_c\gg J,J'$.
A similar nonlinear coupling can be obtained through parametric interactions where a photon of higher energy (the one in the auxiliary cavity in our case) is converted into two photons of half the energy~\cite{Devoret2015}.

As the dissipation rate $\gamma_c$ is the shortest timescale in the problem, the fast motion of the modes in the auxiliary cavity can be neglected, $\partial_t \langle \hat{c}\rangle = 0$~\cite{Zoller2004}.
This procedure allows the replacement of the auxiliary cavity amplitude in the equations of motion of the main system.
Upon the correct choice of $J'$ and $\gamma_c$, the adiabatic elimination of the auxiliary cavity yields the same equations of motion as the two-particle dissipator~(\ref{Eq:Doublon Diss})~\cite{Clerk2015}.
    
\end{document}